\documentclass[11pt,twoside]{article}
\usepackage{asp2010}

\resetcounters

\begin{document}

\title{Evolution of accreting white dwarfs; some of them continue to grow.}
\author{G. Newsham, S. Starrfield, and F. X. Timmes}
\affil{School of Earth and Space Exploration, Arizona State University, Tempe, AZ 85287-
1404}

\begin{abstract}

Novae are cataclysmic variable binary systems in which a white dwarf (WD) primary is accreting material from a low mass companion. The importance of this accretion takes on added significance if the WD can increase its mass to reach the Chandrasekhar limit thus exploding as a Type Ia supernova. In this study we accrete material of Solar composition onto carbon/oxygen (CO) WD’s of 0.70, 1.00 and 1.35$M_{\sun}$ with accretion rates from $1.6 \times 10^{-10} $ to $1.6 \times 10^{-6} M_{\sun}$ yr$^{-1}$. We have utilized the MESA stellar evolution code for our modeling and evolve them for many nova cycles or, in some cases, evolution to a red giant stage. Differing behaviors occur as a function of both the WD mass and the accretion rate.  For the lower WD masses, the models undergo recurrent hydrogen flashes at low accretion rates; for higher accretion rates, steady-burning of hydrogen occurs and eventually gives way to recurrent hydrogen flashes. At the highest accretion rates, these models go through a steady-burning phase but eventually transition into red giants. For the highest WD mass recurrent hydrogen flashes occur at lower accretion rates but for higher rates the models exhibit steady-burning interspersed with helium flashes. We find that for all our models that undergo recurrent hydrogen flashes, as well as the steady-burning models that exhibit helium flashes, the mass of the WD continues to grow toward the Chandrasekhar limit. These results suggest that the accretion of Solar abundance material onto CO WD's in cataclysmic variable systems, the single degenerate scenario, is a viable channel for progenitors of Type Ia supernova explosions. 

\end{abstract}

\section{Introduction}
Cataclysmic variables (CV), including classical and recurrent novae, occur when the accretion of material onto a white dwarf (WD), from a companion star, undergoes a thermonuclear explosion. It has long been thought that such a process could lead to the WD growing to the Chandrasekhar limit resulting in a Type Ia supernova explosion. This is designated the single degenerate scenario. The competing model is the double degenerate scenario that involves the merger or collision of two WDs.

In this paper we investigate the consequence of a WD accreting Solar material from a secondary donor star. Many types of close binary systems with a WD primary have been suggested as the progenitors for Type Ia supernovae; thus we have modeled accretion onto a wide range of WD masses with a wide range of accretion rates. We followed the long-term evolution utilizing the  MESA stellar evolution code. In \S 2 the grid of models is discussed and \S 3 describes the results. In \S 4 we contrast our results with previous work and finish with our conclusions. 

\section{Modeling}

Our modeling was performed with the Modules for Experiments in Stellar Evolution \citep[MESA,][]{pax11,pax13} computer code (version 4798). In our study we used initial WD masses of 0.70, 1.00 and 1.35$M_{\sun}$. All these WDs were initially bare CO cores (C \-- 0.357, O \-- 0.619) prior to accretion commencing. The starting models were at a Solar luminosity. Our mass accretion rates were chosen to be $1.6 \times 10^{-10} $, $1.6 \times 10^{-9} $, $1.6 \times 10^{-8} $ and $1.6 \times 10^{-7}  M_{\sun}$ yr$^{-1}$. Additionally, other accretion rates were used in order to separate different regime behavior when needed. The material being accreted was a Solar mixture \citep{lod03}. Mass loss was triggered when a model reached super-Eddington luminosity. The excess luminosity over Eddington determined the rate of mass loss \citep[see][]{sha02,den13}. All our models were run for many nova cycles or the time required for definitive long-term behavior to become evident.

\section{Results}

\begin{figure}[!ht]
 \epsscale{1.0}
 \plotone{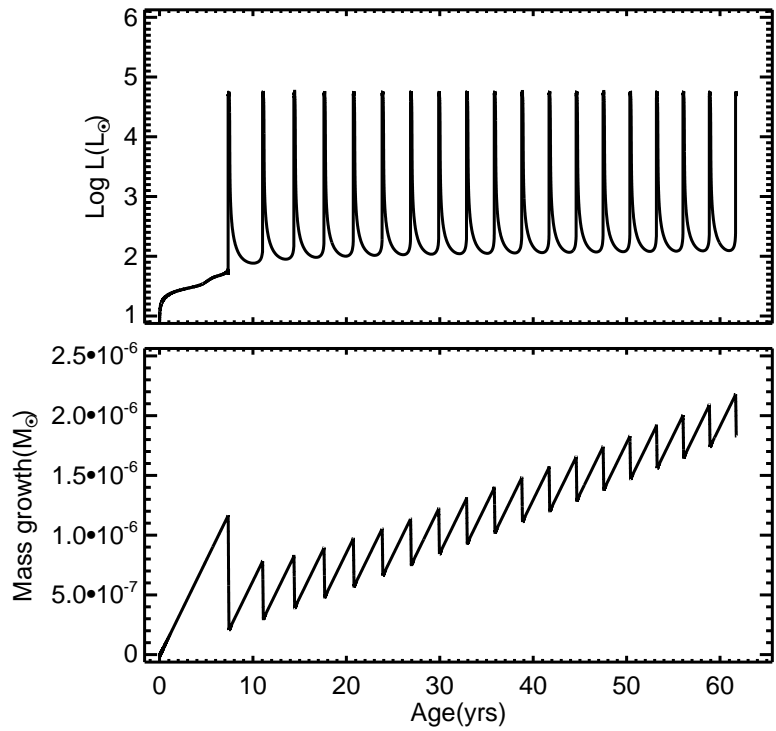}
\caption{Luminosity evolution and mass growth for a 1.35$M_{\sun}$ model accreting at $1.6 \times 10^{-7} M_{\sun}$ yr$^{-1}$ exhibiting recurrent hydrogen flashes. The evident decrease in mass during each flash is caused by mass lost as the outermost layers exceed the Eddington luminosity \citep{sha02}\label{fig:hydrogen}}
\end{figure}

In Figure \ref{fig:hydrogen} (top panel) we show a plot of the surface luminosity versus time for a model exhibiting recurrent hydrogen flashes. In the bottom panel we show the mass of the WD over the same time period. After the initial growth to the first thermonuclear runaway, the model quickly settles into a recurring pattern of flashes in between which mass is accreted. During a flash some of the mass is lost via a super-Eddington wind \citep[see][]{sha02,den13}. The positive slope shows that the WD grows in mass at a rate of $2.96 \times 10^{-8} M_{\sun}$ yr$^{-1}$ for an accretion rate of $1.6 \times 10^{-7} M_{\sun}$ yr$^{-1}$. This represents an efficiency (mass growth versus mass accreted - mass ejected) per cycle of approximately twenty percent.

\begin{figure}[!ht]
\epsscale{1.0} \plotone{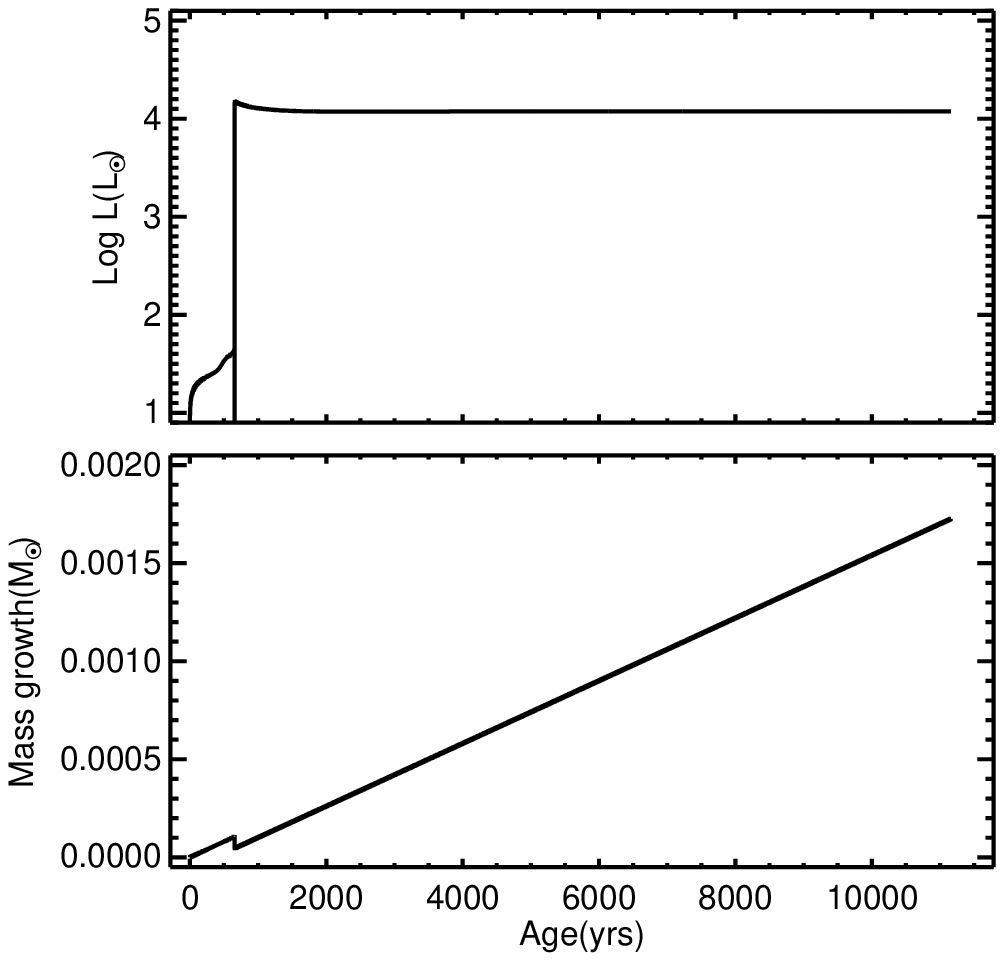} \caption{Luminosity evolution and mass growth for a 0.70$M_{\sun}$ model accreting at $1.6 \times 10^{-7} M_{\sun}$ yr$^{-1}$. There are a few hydrogen flashes followed by steady-burning.\label{fig:steady}}
\end{figure}

In Figure \ref{fig:steady} we show a model/accretion rate combination that exhibits an initial flash and then settles into an extended period of ``steady-burning" where the luminosity is constant. This steady-burning continues until either recurrent flashes, as in Figure \ref{fig:hydrogen} occur, or for higher accretion rates, the model becomes a red giant at which time the evolution is halted. During steady-burning the mass of the WD continues to grow as is shown in the bottom panel of the figure.

The 1.35$M_{\sun}$ WD models exhibit a different behavior at the highest accretion rates. After an initial hydrogen flash the model transitions into a steady-burning phase interrupted by regular helium flashes. During these helium flashes, about half the accreted mass is ejected from the WD. However, the WD continues to grow in mass.

\begin{figure}[!ht]
\epsscale{1.0} \plotone{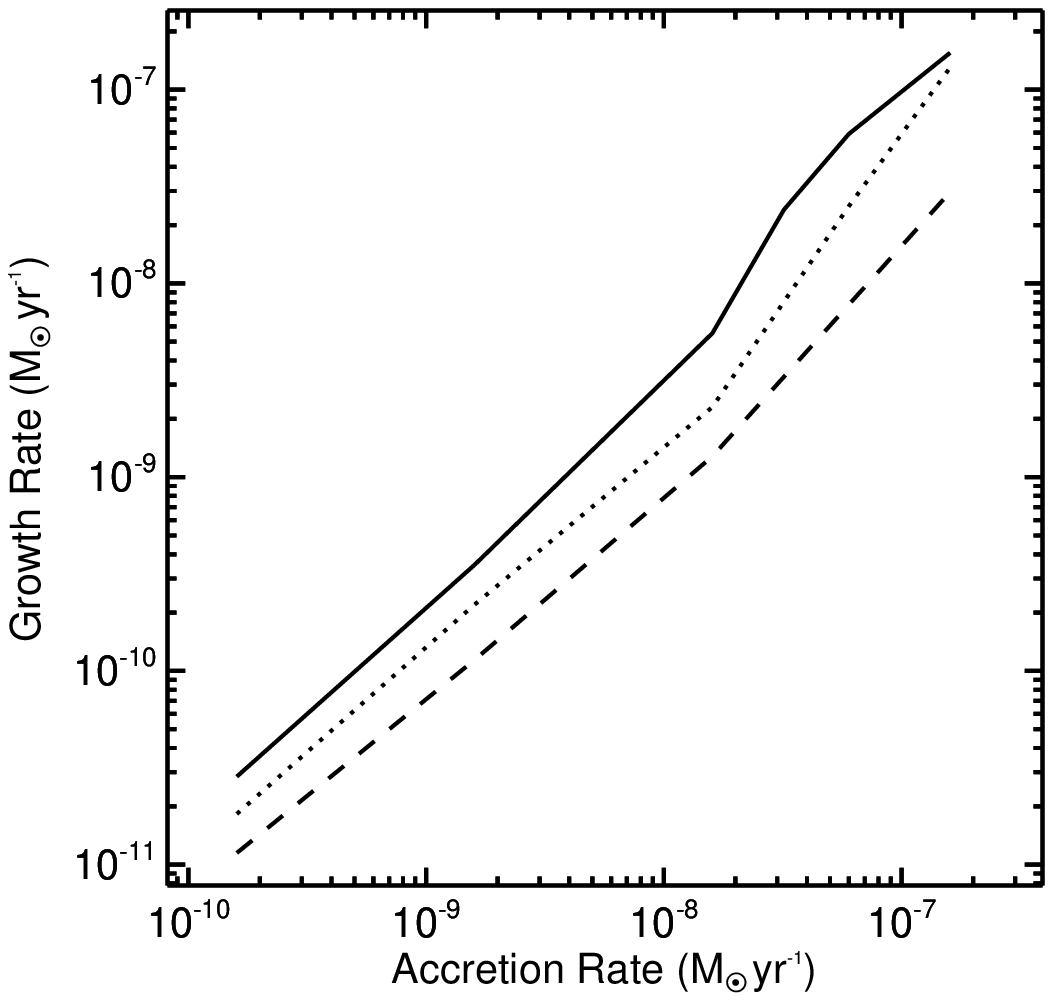} \caption{Growth rate versus the accretion rate for 0.70$M_{\sun}$(solid), 1.00$M_{\sun}$(dotted) and 1.35$M_{\sun}$(dashed).\label{fig:growth}}
\end{figure}

\begin{figure}[!ht]
\epsscale{1.0} \plotone{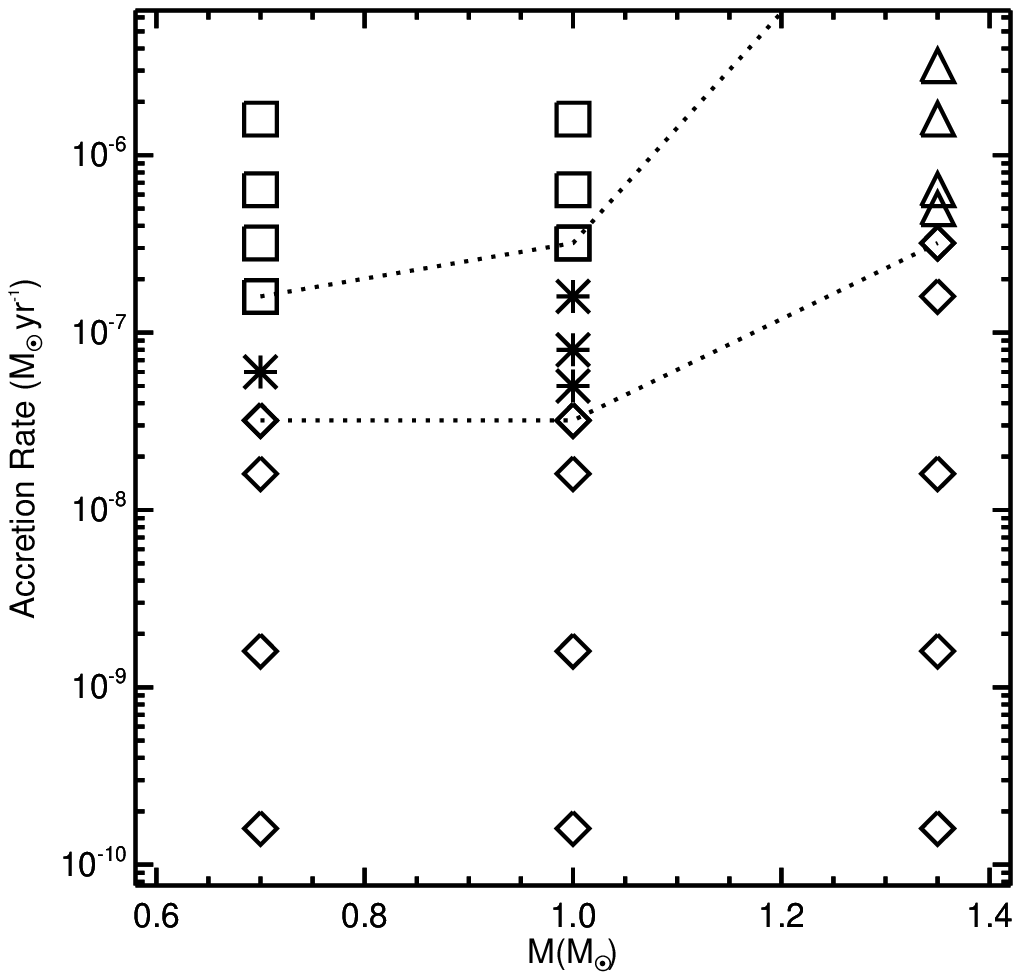} \caption{Accretion rate $\--$ WD mass plane for the range of models we investigated. The symbols indicate models that become red giants (squares), steady-burning followed by hydrogen flashes (stars), steady-burning interrupted by helium flashes (triangles) and recurrent hydrogen flashes (diamonds). Between the dotted lines is the steady-burning region which does not exist at 1.35$M_{\sun}$.\label{fig:nomoto}}
\end{figure}

In Figure \ref{fig:growth} we plot the growth rate versus the accretion rate for all our models that did not terminate in a red giant. The growth rates are much larger for the higher accretion rates at a given WD mass. For a given accretion rate, the smaller WD masses grow in mass at a greater rate and thus at a higher efficiency than the more massive WD models. The 0.70 and 1.00$M_{\sun}$ WDs, at accretion rates greater than approx. $1 \times 10^{-7} M_{\sun}$ yr$^{-1}$, eventually become red giants at which point we end the calculation. However, the 1.35$M_{\sun}$ models, although exhibiting helium flashes, continue their mass growth at the higher accretion rates and never enter a red giant phase.

In Figure \ref{fig:nomoto} we plot the parameter space we investigated in terms of WD mass and accretion rate. We identify the different regimes of behavior in our models and show that, for a large range of accretion rates, our models grow in mass and that it is possible for a model to start off as low as 0.70$M_{\sun}$ and, given sufficient time, will reach 1.35$M_{\sun}$ and beyond. For higher accretion rates, our 0.70 and 1.00$M_{\sun}$ models enter a red giant phase and cease accreting. The 1.35$M_{\sun}$ models, however, never enter such a phase and despite recurrent helium flashes their growth continues unabated.

\section{Discussion}

The different regimes we identify in the accretion of Solar material onto a WD are summarized in Figure \ref{fig:nomoto}. The plot has three main regions. At low accretion rates, the models, for all masses, undergo recurrent hydrogen flashes. At the highest accretion rates for 0.70 and 1.00$M_{\sun}$, the models, after an initial flash enter a regime of steady-burning which eventually leads to the model becoming a red giant. In such a case the accretion ceases and we end the evolution. The third regime, intermediate between the prior two, is where we see steady-burning but the model eventually enters a regime of recurrent hydrogen flashes characteristic of a somewhat lower accretion rate where the flashes occur from the onset of accretion.  

The intermediate behavior is approximately bounded by accretion rates of $5 \times 10^{-8}$ to $1\times 10^{-7} M_{\sun}$ yr$^{-1}$. This is similar to, although slight lower than, the \citet{nom07} results and the theoretical predictions of \citet{she07}. For the 1.35$M_{\sun}$ models, we see a similar regime of steady-burning that is interrupted by regular helium flashes. This occurs only for accretion rates above $5 \times 10^{-7} M_{\sun}$ yr$^{-1}$ and up to the maximum accretion rate we studied of $3.2 \times 10^{-6} M_{\sun}$ yr$^{-1}$.

We find broadly similar results with \citet{cas98} but note some important differences. In their plot highlighting the different regimes (Figure 10) they identified three main regimes. They found, above a certain accretion rate, that a red giant phase ensued without an earlier period of steady-burning as in our results. At low accretion rates, they observed recurrent hydrogen flashes that were more powerful at the lower accretion rate, in agreement with our work. The intermediate steady-burning regime they found, however, shrank in size as a function of WD mass until by 1.00$M_{\sun}$ it ceased to exist. We find that this region is approximately constant in size and in fact increases in the higher mass models, with helium flashes, so that the red giant regime is absent up to the upper limit of accretion rate in our study. It should be noted that they only investigated lower mass WDs and included data from \citet{liv89} for 1.00$M_{\sun}$ WDs.

\citet{cas98} concluded that, for the high accretion rate systems, the red giant phase resulted in a common envelope that would eject most if not all of the matter donated by the secondary star. Thus these systems were not progenitors of Type Ia supernovae. They also concluded that at the lowest accretion rates, responsible for the strongest hydrogen flashes, produced a classical novae explosion and that these systems could not be progenitors of Type Ia supernovae.  For the models between these limits they were limited (by their computing capability) to follow only a few hydrogen flashes. It should be noted, however, that for one particular case they followed a model until a large helium flash occurred. They surmised that such a helium flash would curtail the growth of the WD and thus such systems were also precluded as being the progenitors of Type Ia supernovae.

We found no helium flashes in our low mass models for any accretion rate. For the steady-burning models they found helium flashes occurred both for low mass models (0.516 and 0.6$M_{\sun}$) and the higher mass 0.8$M_{\sun}$ model. They stated that such flashes would result in envelope expansion beyond the Roche lobe of the WD and that the envelope is lost from the system in a similar manner to the WD becoming a red giant.

In contrast, the only helium flashes we find are for steady-burning models at 1.35$M_{\sun}$. One particular case was for a 1.35$M_{\sun}$ WD accreting at a rate of $6.4 \times 10^{-7} M_{\sun}$ yr$^{-1}$. After approximately seventy-five years of steady-burning the system underwent a helium flash in which $2.5 \times 10^{-5} M_{\sun}$ was ejected. The radius of the WD peaked at  0.056$R_{\sun}$ and the peak surface velocity was 10 km s$^{-1}$. If this WD were in a CV system with a typical companion that had a mass of 0.7$M_{\sun}$ and a radius of 0.75$R_{\sun}$ the secondary would fill its Roche lobe if the system had a semi-major axis of 2.32$R_{\sun}$ and period of 6.9 hours. This gives a Roche lobe radius for the WD primary of 1.01$R_{\sun}$. However, our WD model never exceeded more than about six percent of this during the helium flash. We also note that the peak velocity of 10 km s$^{-1}$ is only about 1/30th of the escape velocity of the WD which is 2700 km s$^{-1}$. We conclude such helium flashes do not curtail the growth of our models due to Roche lobe overflow or complete envelope ejection. 

The reason for our WD helium flashes being less powerful is that we model our mass loss as a super-Eddington wind which prevents the atmosphere from becoming extended. Thus a helium flash is not the dynamic mass ejection that occurs using a mass loss prescription like \citet{pri78, pri79}. In such simulations very large radii result from an initial thermonuclear runaway and further evolution is curtailed \citep[see][]{sta12b,sta12a}. More sophisticated treatments of optically thick wind mass loss \citep[see][]{kov98,ida12} tend to produce greater mass loss for a given accretion rate and thus correspondingly less efficient growth of the WD primary. In \citet{ida12} modeling, steady-burning is quenched and is replaced by a series of recurrent hydrogen flashes, which build up a helium layer until a large helium flash occurs removing all of the accreted mass. However, at lower accretion rates their models grow in mass just like ours though less efficiently.

From our Figure \ref{fig:growth} we can estimate the timescale for the growth of systems to reach the Chandrasekhar mass and thus be candidate progenitors for Type Ia supernovae. We discard the models that become red giants and the lowest accretion rate models with the most powerful hydrogen flashes (accretion rate less than $10^{-8} M_{\sun}$ yr$^{-1}$), assuming that they mix in some fashion and become classical novae. The time for the 0.70$M_{\sun}$ model to attain the Chandrasekhar limit is $8 \times 10^{6}$ to $4 \times 10^{8}$ years. The 1.00$M_{\sun}$ model takes $6 \times 10^{6}$ to $3 \times 10^{8}$ years and the 1.35$M_{\sun}$ case requires $3 \times 10^{6}$ to $1 \times 10^{8}$ years. The shortest and longest timescales are seen to vary by two orders of magnitude.

\section{Summary}

We conclude that the single degenerate scenario for Type Ia supernova production is viable as long as the accreted material does not mix with the core material. Not all combinations of WD mass/accretion rates lead to continued mass growth of the WD primary but some do allow accretion to continue unabated. However, at lower WD masses, high accretion rates lead to red giant formation of the WD, which through Roche lobe overflow and a subsequent common envelope phase eject mass from the system. Though we did not investigate very low accretion rates, where diffusion mixing becomes important, it is suspected that the more powerful hydrogen flashes would eject enough mass to prevent growth to the Chandrasekhar limit.

Our models that undergo recurrent hydrogen flashes or steady-burning followed by either hydrogen or helium flashes continue to grow in mass. We find that the helium flashes are not powerful enough to curtail the mass growth. The first point is in agreement with that of \citet{ida12} though the second point, in regard to helium flashes, is at odds with their work. This is most likely due to the different approach to mass loss; their optically thick wind model as opposed to our super-Eddington mass loss prescription. 

We find that for our WD models that grow in mass; the timescale for reaching the Chandrasekhar limit varies by two orders of magnitude, dependent upon initial mass and accretion rate, from $3\times10^{6}$ to $4\times10^{8}$ years.
These results suggest that Solar type material accreted onto a CO WD is a channel for Type Ia supernovae production.

\acknowledgements We would like to thank the many people who attended the Capetown meeting for their interest in these results. G. Newsham, F. Timmes and S. Starrfield acknowledge partial support from NSF grant AST10-07977 to Arizona State University. S. Starrfield also acknowledges partial support from other NSF and NASA grants to Arizona State University.

\bibliography{newsham}

\end{document}